\documentclass[12pt,preprint]{aastex}
\newif\ifjournal\journalfalse

\usepackage{graphicx}
\usepackage{bm}
\renewcommand{\vec}[1]{\boldsymbol{#1}}

\newcommand{\rot}{\nabla \times}

\def\jgr{{\itshape J. Geophys. Res.} }
\def\apj{{\itshape Astrophys. J.} }

\def\prl{{\itshape Phys. Rev. Lett.} }
\def\pop{{\itshape Phys. Plasmas} }
\def\aap{{\itshape Astron. Astrophys.} }

\def\mnras{{\itshape Monthly Notices of the RAS} }

\shorttitle{Self-Regulation of Relativistic Reconnection}
\shortauthors{Zenitani \& Hesse}

\begin{document}

\title{Self-regulation of the reconnecting current layer\\
in relativistic pair plasma reconnection}

\author{S. Zenitani and M. Hesse}
\affil{
NASA Goddard Space Flight Center, Greenbelt, MD 20771
}
\date{Received 2008 Mar 28; accepted 2008 May 28}

\begin{abstract}
We investigate properties of the reconnecting current layer
in relativistic pair plasma reconnection.
We found that
the current layer self-regulates its thickness
when the current layer runs out current carriers,
and so relativistic reconnection retains a fast reconnection rate.
Constructing a steady state Sweet-Parker model,
we discuss conditions for the current sheet expansion.
Based on the energy argument,
we conclude that the incompressible assumption is invalid
in relativistic Sweet-Parker reconnection.
The guide field cases are more incompressible than
the anti-parallel cases, and
we find a more significant current sheet expansion.
\end{abstract}

\keywords{magnetic fields --- plasmas --- relativity}

\section{INTRODUCTION}

Magnetic reconnection \citep{vas75,biskamp,birn07} is the driver of
explosive events in space plasmas and laboratory experiments. 
Although it has extensively been studied ever since the 1950s,
its detailed mechanism still remains unclear.
The complexity of the reconnection problem is that
small-scale physics in and around
the reconnecting $X$-type region (or the diffusion region)
can drastically change the system's global evolution.

Magnetic reconnection has drawn attention
in various high-energy astrophysical applications
\citep{coro90,drs02,lyut03a},
where a relativistic extension of magnetic reconnection
in an electron-positron pair plasma ($e^\pm$) is considered.
However, the physical mechanism of relativistic magnetic reconnection,
as well that of the well-studied nonrelativistic reconnection,
is far from being understood.
For example, only a few researchers have developed 
fundamental models of relativistic steady state reconnection.
\citet{bf94b} predicted that
relativistic magnetic reconnection facilitates
faster energy conversion
than nonrelativistic reconnection,
due to the Lorentz contraction in the outflow region.
\citet{lyutikov03} followed them, but
\citet{lyu05} claimed that reconnection inflow cannot be so fast.
He further worked on
the generalization of relativistic Petschek reconnection
and argued that the guide field (out-of-plane field) opposes
energy conversion into the plasma energy.

On the other hand,
self-consistent simulations have revealed
important aspects of relativistic magnetic reconnection.
Several authors \citep{zeni01,zeni05b,zeni07,zeni08,claus04,bessho07} carried out
particle-in-cell (PIC) simulations of
relativistic pair plasma reconnection,
and they found that powerful particle acceleration occurs
inside/near the diffusion region. 
The main accelerator is the reconnection electric field,
which becomes strong enough to sustain relativistic outflow.
When the system contains a guide field,
relativistic magnetic reconnection also involves particle acceleration
by the reconnection electric field
\citep{zeni05b,zeni08,karl08}.
It seems that particle acceleration by the reconnection electric field 
is a common feature of relativistic magnetic reconnection.
Furthermore, carrying out a relativistic magnetohydrodynamic (MHD) simulation,
\citet{naoyuki06} presented a Petschek structure in mildly-relativistic regime.
Note that they assumed spatially limited resistivity,
representing the physics inside the diffusion region.

In an ion-electron plasma,
it is known that a thin current layer
appears near the $X$-point during magnetic reconnection
\citep{hori94}.
The main current carriers are electrons,
and the thin electron current layer is a key region to 
understand the reconnection physics.
So, significant efforts have been paid to
the evolution of the electron diffusion region
\citep{hesse98,hesse99}.
The general consequence will be that
the current layer becomes thin as reconnection develops,
typically into the electron meandering width \citep{hori94}.
On the other hand, there is an ongoing discussion on
the length of the current layer \citep{keizo06,dau06,shay07},
whose extension may regulate the reconnection rate
by changing the aspect ratio of the ``diffusion'' region.

In a relativistic pair plasma,
little attention has been paid to
the detailed structure of the diffusion region,
although it is a key region to drive magnetic reconnection
and to accelerate high-energy particles.
Recently, \citet{hesse07} investigated
the composition of the reconnection electric field,
and they showed that the off-diagonal parts of the pressure tensor are important,
similar to the nonrelativistic cases \citep{hesse99,hesse04,bessho07}.
However, a physical interpretation of the off-diagonal terms
has not yet been established
as well as the terms
in a true decomposition of the relativistic pressure tensor.
In the present paper
we investigate the properties of the diffusion region
in relativistic pair plasma reconnection
from another viewpoint:
the temporal development of the reconnecting current layer. 
The unexpected new result is that
the out-of-plane current layer adjusts itself by expanding its thickness.
When the plasma drift velocity reaches an upper limit,
typically on the order of the light speed,
the current layer runs out current carriers.
Then, the reconnection electric field becomes stronger
to satisfy Amp\`{e}re's law for the field reversal current.
Consequently, the non-MHD diffusion region and the current layer become wider.
Since it enhances the reconnection electric field and
since it improves the aspect ratio of the dissipation region,
relativistic reconnection retains a fast reconnection rate.
A brief discussion of the steady state energy flow
provides some insight into this process.
We cannot employ the incompressible assumption in
relativistic Sweet-Parker conditions any more.
The guide field brings incompressibility, and therefore,
we find a more significant current sheet expansion.

The paper consists of the following sections.
In \S 2 we describe our simulation setup and
briefly overview the system evolution.
In \S 3 we extended our interpretation
by using a steady-state Sweet-Parker model.
In \S 4 we discuss the guide field extension.
Finally, \S 5 contains discussion and the summary.

\section{SIMULATION AND RESULTS}

We carry out two-dimensional PIC simulations
in the $x$-$z$ plane.
As an initial current sheet configuration,
we employ a relativistic extension of the Harris model.
The magnetic field, plasma density, and
plasma distribution functions are described by
$\vec{B} = B_0 \tanh(z/L) \hat{\vec{x}} + B_G \hat{\vec{y}}$,
$d(z) = (\gamma_{\beta} n_0) \cosh^{-2} (z/L)$ and
$f_{s} \propto d(z) \exp[-\gamma_\beta\{\varepsilon - \beta_s u_y\} / T ] $.
In the above equations,
$B_0$ is the magnitude of antiparallel magnetic field,
$B_G$ is the out-of-plane magnetic field,
$L$ is the typical thickness of the current sheet,
$n_0$ is the proper number density of plasmas in the current sheet,
the subscript $s$ denotes the species
(\textit{p} for positrons, \textit{e} for electrons),
$\beta_{p} = -\beta_{e} = \beta$ is the dimensionless drift velocity,
$\gamma_\beta$ is the Lorentz factor for $\beta$ ($\gamma_\beta = [1-\beta^2]^{-1/2}$),
$\varepsilon$ is the particle energy,
$\vec{u}$ is the relativistic four-velocity of
$\vec{u}= [1-(v/c)^2]^{-1/2} \vec{v}$,
and $T$ is the proper temperature including the Boltzmann constant.
We set $T = mc^2$ and $\beta = 0.3$, respectively.
In addition, a uniform background plasma is added to the system.
Its number density and temperature are
$n_{bg}/(\gamma_{\beta}n_0)=5\%$ and $T_{bg}/mc^2=0.1$, respectively.
This background temperature is cold enough that
the background plasma energy density is
$\sim 1.2 n_{bg}mc^2$ per species,
including the rest-mass energy.
The plasma satisfies the pressure balance condition
$B_0^2/8\pi=2n_0T$ in this equilibrium.
The Debye length is
$[ T / (4\pi \gamma^2_{\beta} n_0 q^2)]^{1/2} = 0.3 L$,
where $q$ is the charge.
The typical Larmor radius is $c(qB_0/\gamma mc)^{-1} \sim 0.2 L$.

The system consists of $1536 (x) \times 768 (z)$ cells,
and the typical scale of the current sheet $L$ is set to 10 cells.
We consider periodic boundaries in the $x$-direction,
and the boundaries are located at $x = \pm 76.8 L$.
We also consider periodic boundaries in the $z$-direction,
but we divide the simulation domain into two subdomains.
The first half (the main simulation domain) is located in
$-19.2 \le z/L \le 19.2$, and the other half (the subsimulation domain) is
located in $19.2 \le z/L \le 57.6$.
In the subdomain,
we assume an oppositely directed Harris model 
(e.g. $\vec{B} = -B_0 \tanh[z/L-38.4]\vec{\hat{x}}$),
so that physical properties in the sub domain are smoothly connected to
those in the main domain.
We use $7.6 \times 10^7$ superparticles in this simulation.
One cell contains $6.4 \times 10^2$ particles
at the center of the current sheet. 
The detailed parameters are presented in Table \ref{table}.
During the very early stage of the simulation,
we set a small driving force to trigger magnetic reconnection
near the center of the simulation domain.
This trigger field smoothly varies in time,
and it is applied until $t \sim 12\tau_c$,
where $\tau_c=L/c$ is the light transit time,
while we discuss the physics of reconnection
in the late stage of $30\tau_c \sim 90\tau_c$.
The total energy is conserved within an error of $0.1\%$
throughout the simulation run, after the initial trigger force vanishes.
In addition, we carry out four runs with different parameters
of $n_{bg}/(\gamma_{\beta}n_0)$ and $B_{G}/B_{0}$.
Their parameters are shown in Table \ref{table}.
These configurations are similar to previous simulations
\citep{zeni07,zeni08}.

\begin{figure}[htbp]
\begin{center}
\ifjournal
\includegraphics[width={\columnwidth},clip]{f1.eps}
\else
\includegraphics[width={\columnwidth},clip]{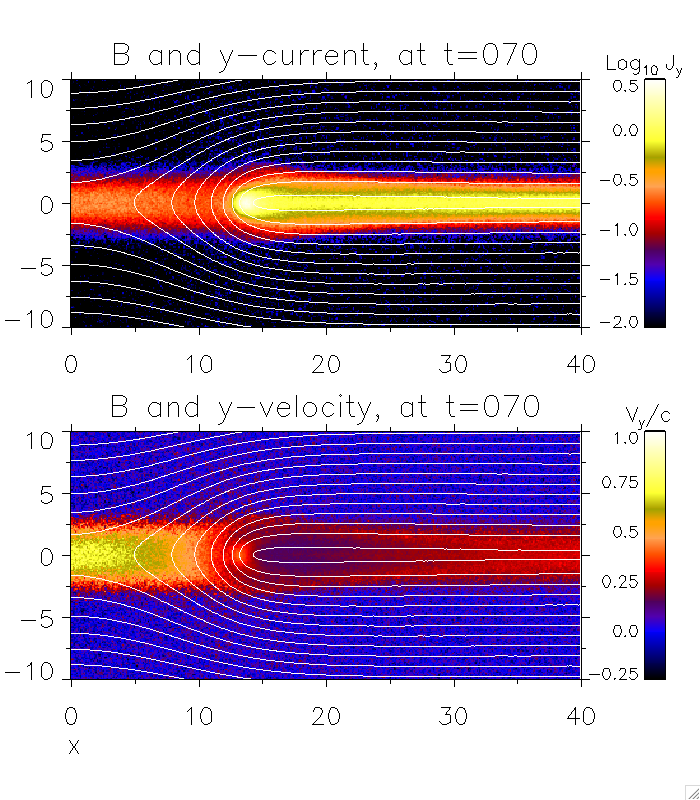}
\fi
\caption{Magnetic field lines and \textit{y}-component of
(\textit{a}) the current density and
(\textit{b}) the average velocity at $t/\tau_c=70$ in run 1.
\label{fig:snap}}
\end{center}
\end{figure}

Here we describe the evolution of the first run (run 1).
After the initial impact disappears,
magnetic field lines start to reconnect around the center
of the main simulation domain.
The panels of Figure \ref{fig:snap} present
the snapshots of the reconnecting region at $t/\tau_c=70$.
The $X$-point is located at the left border ($x=0$),
while the simulation domain is $-76.8<x/L<76.8$.
The top panel shows magnetic field lines and
the $y$-current (out-of-plane component of the current) density.
The bottom panel shows magnetic field lines and
the $y$-component of the plasma average velocity.
Notice that the $X$-point is located around $x \sim 0$.
At this stage, magnetic reconnection is well developed:
the plasma outflow velocity ($\langle v_x \rangle$) is
up to $\sim 0.7c$ around the outflow region.

\begin{figure}[htbp]
\begin{center}
\ifjournal
\includegraphics[width={\columnwidth},clip]{f2.eps}
\else
\includegraphics[width={\columnwidth},clip]{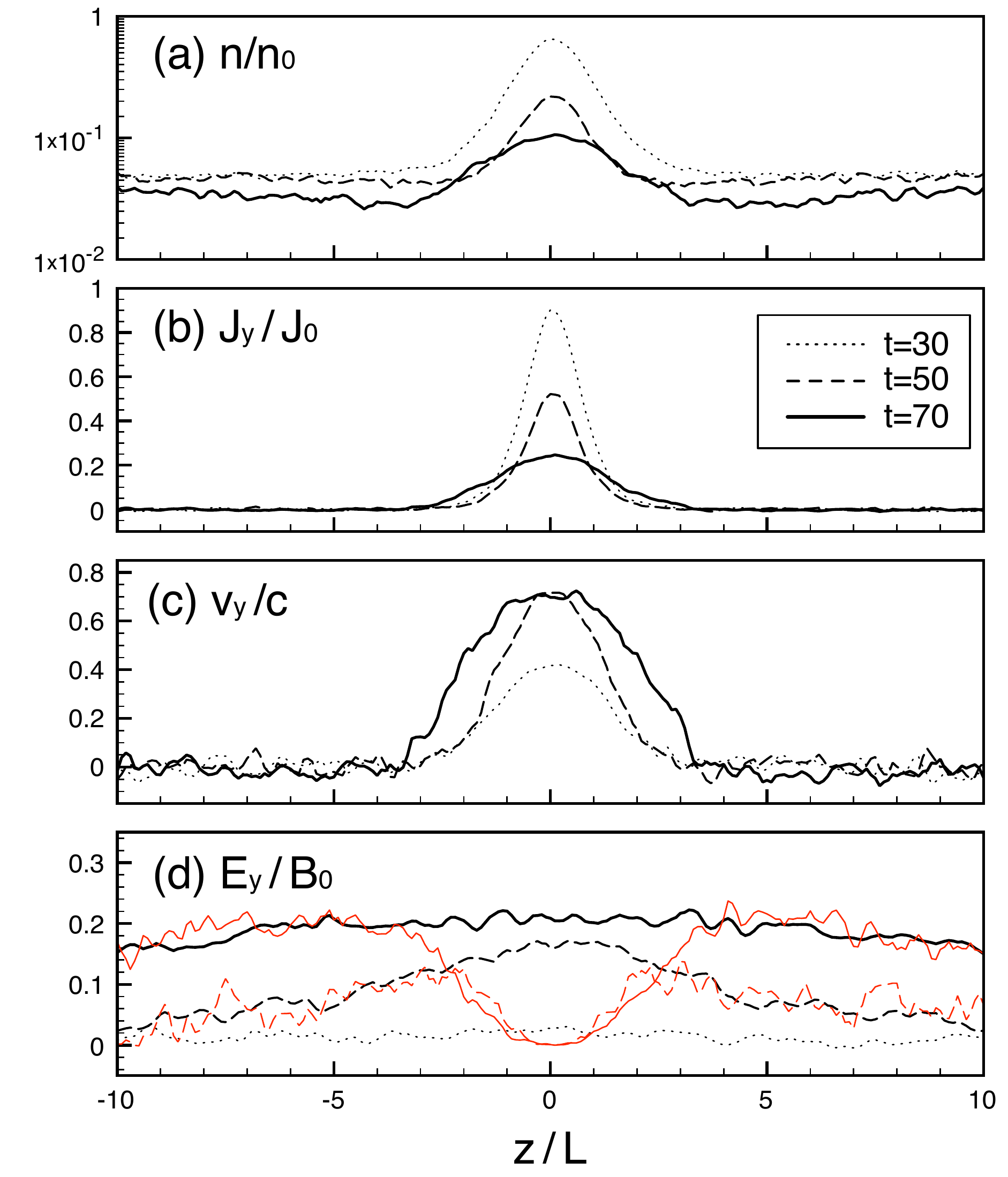}
\fi
\caption{Physical properties along the inflow region
at $x/L=0$ in run 1:
(\textit{a}) normalized plasma density,
(\textit{b}) $y$-component of the electric current,
(\textit{c}) $y$-component of the plasma average velocity, and
(\textit{d}) reconnection electric field $E_y$.
The red lines show the relevant $(-\vec{v}\times\vec{B})_y$ electric fields.
\label{fig:plot}}
\end{center}
\end{figure}

Panels in Figure \ref{fig:plot} present
the plasma number density $2n$, the current density $J_y$,
average drift velocity $\langle v_y\rangle$,
and reconnection electric field $E_y$
along the inflow line $x=0$.
These values are averaged over $-1/4 \le x/L \le 1/4$
in order to reduce noise.
The $y$-current is carried by the drift motion of
positrons ($+y$-direction) and electrons ($-y$-direction)
inside the current sheet, $J_y \sim 2 d(z)\langle v_y\rangle$.
We present three characteristic stages at $t/\tau_c=30, 50,$ and $70$.
Shortly before $t/\tau_c=30$,
reconnection is about to begin:
$\langle v_y\rangle$ and $E_y$
start to increase around the $X$-type region,
while plasma density starts to decrease there,
because plasmas flow away into the reconnection outflow region.
At $t/\tau_c=50$,
the reconnecting current layer looks ``thin''
around the reconnecting region,
and then the plasma inflow and outflow structures are developed.
The $\langle v_y \rangle$ goes up to $\sim 0.7c$ around the center.

\begin{figure}[htbp]
\begin{center}
\ifjournal
\includegraphics[width={\columnwidth},clip]{f3a.eps}
\else
\includegraphics[width={\columnwidth},clip]{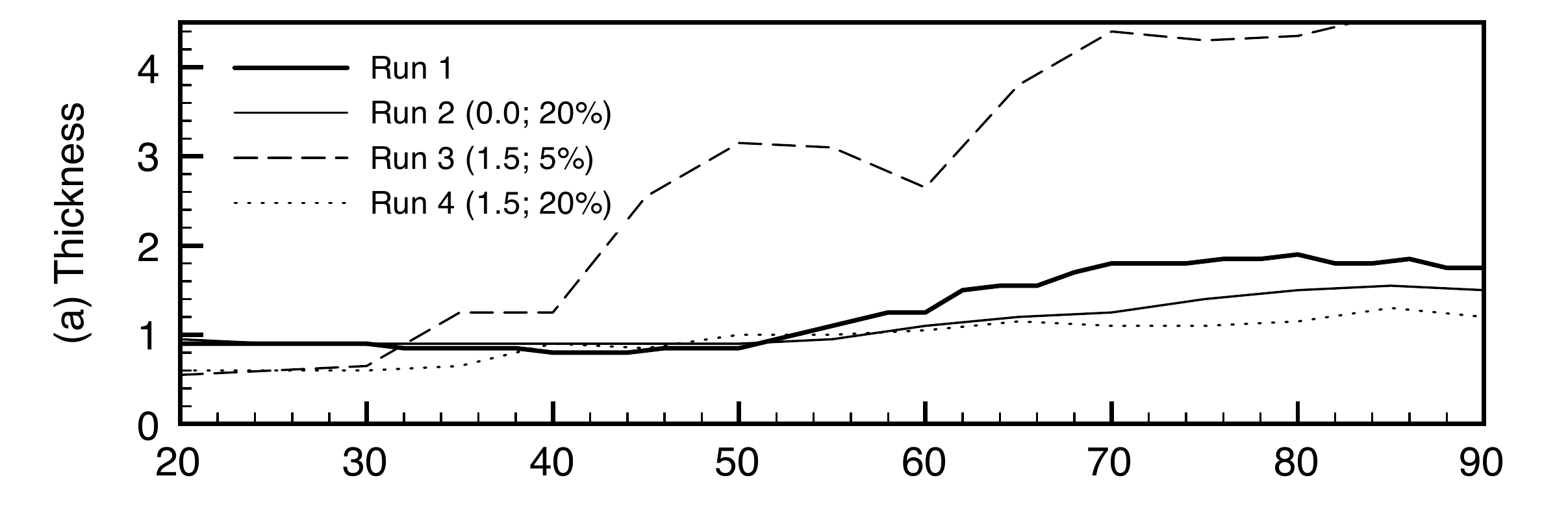}
\fi
\ifjournal
\includegraphics[width={\columnwidth},clip]{f3b.eps}
\else
\includegraphics[width={\columnwidth},clip]{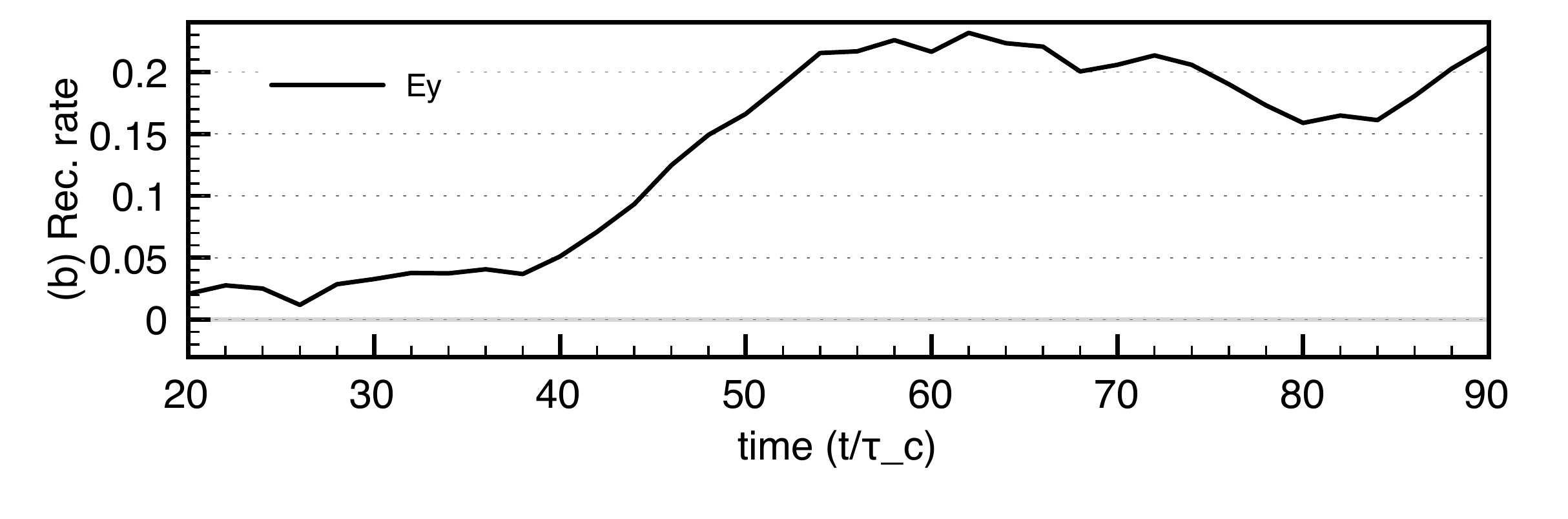}
\fi
\caption{
(\textit{a})
Temporal evolution of the current sheet thickness at $x/L=0$,
measured by the scale height of the current density
($\log J_y$).
(\textit{b})
Temporal evolution of the reconnection rate
($E_y/B_0$) in run 1.
\label{fig:time}}
\end{center}
\end{figure}

After $t/\tau_c=50$,
the current layer shows an unexpected evolution
which is not explained by present reconnection theories.
Previous studies show current layers,
which are substantially thinner than the initial current sheet.
However, in this case,
the current sheet starts to expand.
The change may look small in Figure \ref{fig:snap}, but
we can clearly recognize the expansion signature
in Figures \ref{fig:plot}\textit{b} and \ref{fig:plot}\textit{c}.
The plasma velocity $\langle v_y \rangle$ and
the current density $J_y$ increase
outside the initial current sheet.
In fact, $J_y$ outside the current sheet
is stronger than in the initial state,
which was almost zero.
Figure \ref{fig:time}\textit{a} (\textit{black line}) presents
the time evolution of the current sheet thickness
measured by the scale height or the relative amplitude
of the current density $|J_y|$.
During the preflight stage,
the current sheet becomes slightly thinner
than the initial Harris sheet thickness,
but it starts to expand,
and finally it becomes $\sim 1.9$ times wider after $t/\tau_c=70$.

Figure \ref{fig:time}\textit{b}
shows time evolution of the reconnection rate, 
represented by the electric field $E_y$ at the $X$-point.
In this two-dimensional configuration,
the reconnection rate ${E_y}/{B_0}$ immediately indicates
the efficiency of magnetic energy conversion.
The rate is often normalized by the reconnection outflow speed
in nonrelativistic studies,
but we normalize the rate by $c$,
since the outflow is on the order of $\sim 0.7c$.
So, although this rate may be slightly underestimated,
we obtain a relatively fast reconnection rate of $0.15$-$0.2$.
This rate is faster than
the well-known nonrelativistic reconnection rates of $\sim 0.1$,
due to the $E_y$-enhancement along with the current sheet expansion.

Next, let us interpret the physics of
the current sheet expansion.
In this simulation, we use a large density ratio
of Harris sheet plasmas to the background plasmas,
$n_{bg}/(\gamma_{\beta}n_0)=5\%$.
Once reconnection starts,
dense Harris sheet plasmas escape into the outflow region,
while only less dense background plasma enters the reconnecting region.
Therefore, the plasma density continues to decrease
as seen in Figure \ref{fig:plot}\textit{a}.
The density finally decreases to
about the background density.
Consequently, the system starts to run out
the $y$-current $J_y$
for the field reversal around the diffusion region.
Recall the restriction of Amp\`{e}re's law:
the system needs the out-of-plane current $J_y$,
\begin{equation}
\frac{\partial B_x}{\partial z} \sim
(\rot \vec{B})_y = \frac{4\pi}{c}{J}_y
+ \frac{1}{c}\frac{\partial E_y}{\partial t}
\end{equation}
By some mechanism, the system needs to maintain the field reversal.

The first workaround is
to enhance the plasma current $J_y=2d_{cs}\langle v_y\rangle$,
where $d_{cs}$ is the number density in the current sheet.
In the nonrelativistic regime,
the plasma current $J_y$ is simply enhanced by
particle acceleration into the $\pm y$-directions.
The displacement current works indirectly,
because the enhanced electric field $E_y$ accelerates particles.
In fact, at $t/\tau_c \lesssim 50$,
the reconnection electric field $E_y$ and 
the plasma $y$-velocity $\langle v_y\rangle$ are significant
only around the center ($z\sim 0$;
Figs. \ref{fig:plot}\textit{c} and \ref{fig:plot}\textit{d}). 
However, in a relativistic plasma,
the plasma current has an upper limit of $|\vec{J}| < 2d_{cs}c$.
When the velocity becomes on the order of $c$,
plasmas cannot carry any more current. 

The next workaround is to directly use
the displacement current $(1/c)(\partial\vec{E}/\partial t)$.
Indeed, in this case,
the plasma current explains only 80\% of the required current
and the displacement current accounts for the other 20\% around
the turning phase of $t/\tau_c \sim 50$.
However, the system cannot use
the displacement current to maintain
a steady or quasi-steady structure.
This workaround is temporal,
and then the system goes into the other phase
around $t/\tau_c \sim 50$.

Because of the effects of the displacement current term,
the current layer starts to expand by the derivative
of the generated electric field.
Because of the displacement current,
the electric field $E_y$ becomes even stronger
across the current layer.
This reduces magnetic fields outside the current layer,
\begin{equation}
\frac{\partial B_x}{\partial t} = -(c \rot \vec{E})_x \sim c \frac{\partial E_y}{\partial z}.
\end{equation}
Note that $B_x>0$ and $(\partial E_y/\partial z)<0$
on the upper side ($z>0$),
while $B_x<0$ and $(\partial E_y/\partial z)>0$
on the lower side ($z<0$).
Consequently, the MHD frozen-in condition
breaks down over the wider spatial region around the neutral sheet.
The red lines in Figure \ref{fig:plot}\textit{d} present
$(-\vec{v}\times \vec{B})_y$ at the relevant stages.
We recognize that the non-MHD diffusion region
of $E_y \ne (-\vec{v}\times \vec{B})_y$ becomes wider
during $t/\tau_c = 50 \rightarrow 70$.
Obviously, the non-MHD region corresponds to
the plasma drift region of $\langle v_y \rangle \gtrsim 0$.
Since plasmas are free from the frozen-in restriction
within the diffusion region,
they are accelerated into the $\pm y$ directions
by the enhanced electric field $E_y$,
and then they start to carry the $y$-current.
At $t/\tau_c = 70$
we can recognize that
the outer regions around $z/L\sim\pm 2$
carry the $y$-current
(Figs. \ref{fig:plot}\textit{b} and \ref{fig:plot}\textit{c}) and
become a part of the diffusion region (Fig. \ref{fig:plot}\textit{d}).
The enlarged diffusion region involves more plasmas
outside the initial current sheet,
and then they are responsible for the required $J_y$ for the field reversal.

In Figure \ref{fig:time}
we can compare the trend of
the current sheet thickness and the reconnection electric field.
The reconnection electric rate $E_y$ increases first
and then it stops before $t/\tau_c \sim 60$.
The current sheet thickness slightly delays
to the reconnection electric field,
and it starts to expand.
The current sheet remains nearly constant after $t/\tau_c=70$-$80$.

\section{ANALYTIC THEORY}

\begin{figure}[htbp]
\ifjournal
\includegraphics[width={\columnwidth},clip]{f4.eps}
\else
\includegraphics[width={\columnwidth},clip]{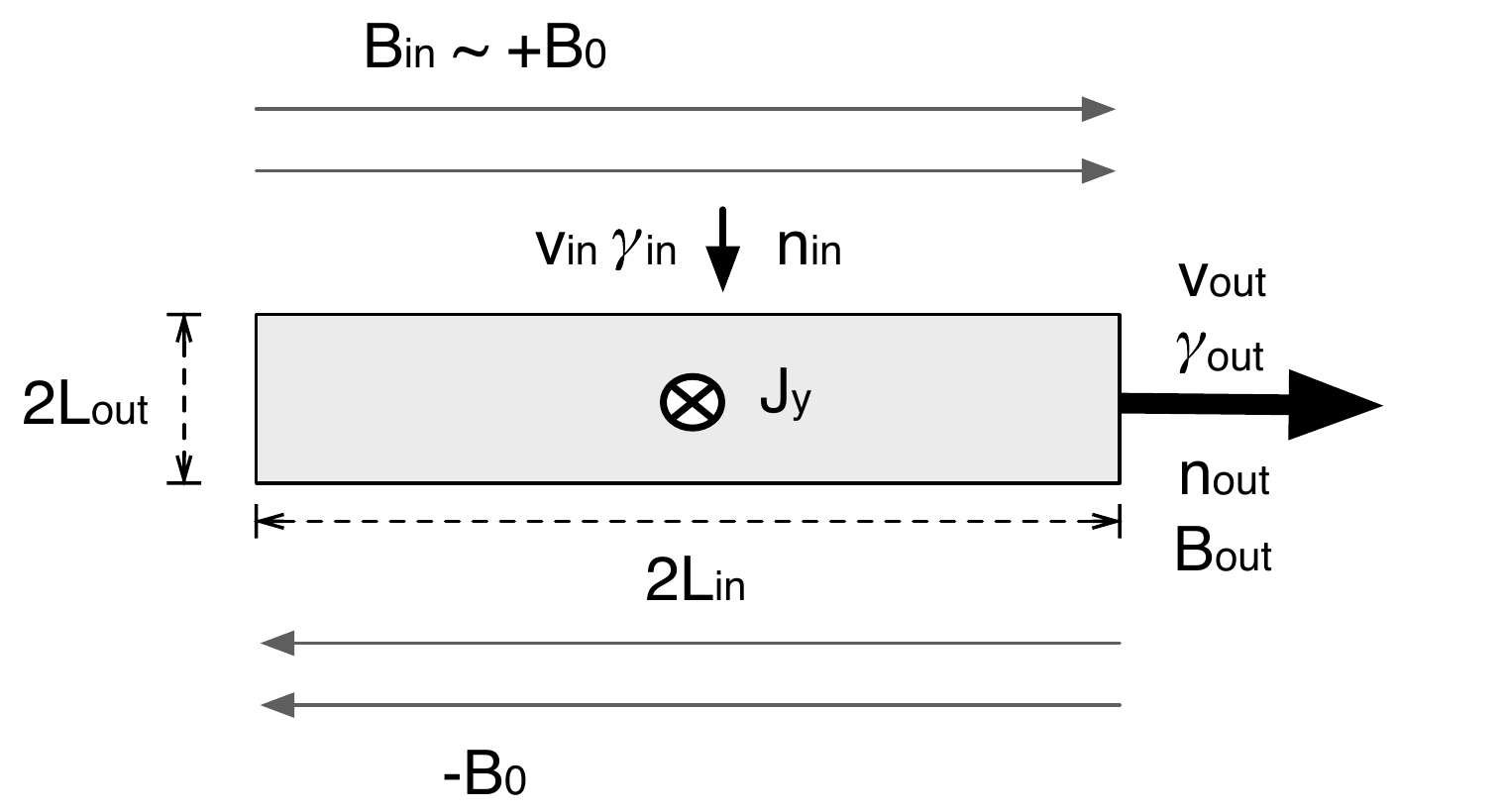}
\fi
\caption{Simple model for a Sweet-Parker current sheet.
Inflow properties and outflow properties are 
represented by subscripts ``in'' and ``out,''
respectively.
\label{fig:model}}
\end{figure}

In order to understand the current sheet expansion,
we consider a steady state Sweet-Parker current sheet model
as presented in Figure \ref{fig:model}.
This model is too simple to discuss the detailed properties of simulation runs,
but sufficient to understand the physics.
In addition, as long as we consider a mildly relativistic regime of our scope,
the current layer has a simple planar structure
around the entire reconnecting region (Fig. \ref{fig:snap}).
In Figure \ref{fig:model}
the subscripts ``in" and ``out" denote
the physical properties
in the inflow region and in the outflow region, respectively.
The current sheet width and height are
$2L_{in}$ and $2L_{out}$. 
In the steady state,
conservation of magnetic flux,
the number density continuity
and energy budget
in the model current sheet are written as
\begin{eqnarray}
\label{eq:flux}
B_{in}v_{in} &=& B_{out}v_{out}\\
\label{eq:continuity}
\gamma_{in} n_{in} L_{in} v_{in} &=& \gamma_{out} n_{out} L_{out} v_{out} \\
\label{eq:eflow}
\Big( 2\gamma^2_{in} w_{in} + \frac{B_{in}^2}{4\pi} \Big)
L_{in} v_{in}
&=&
\Big( 2\gamma^2_{out} w_{out} + \frac{B_{out}^2}{4\pi} \Big)
L_{out} v_{out}
,
\end{eqnarray}
where $w=e+p$ is the plasma enthalpy density,
the sum of the plasma internal energy ($e$) and the proper pressure ($p$).
We introduce the factor of 2 in equation \ref{eq:eflow}
so that we can consider both electrons and positrons.
In the $x$-$z$ plane,
we can assume that the two species move together.
By using the ratio of Poynting flux to the particle energy flux
\begin{eqnarray}
\label{eq:sigma}
\sigma = \frac{B^2}{4\pi (2\gamma^2 w)},
\end{eqnarray}
equation \ref{eq:eflow} can also be written as
\begin{eqnarray}
\label{eq:eflow2}
\Big( \frac{ 1+\sigma_{in} }{ \sigma_{in} } \Big) \frac{B_{in}^2}{4\pi}
L_{in} v_{in}
&=&
\Big( 2\gamma^2_{out} w_{out} + \frac{B_{out}^2}{4\pi} \Big)
L_{out} v_{out}
.
\end{eqnarray}
Note that the relativistic Alfv\'{e}n velocity is $V_A = [\sigma/(1 + \sigma )]^{1/2}c$
and the relevant Lorentz factor is $\gamma_A = (1 + \sigma)^{1/2}$.
In a cold plasma limit,
$\sigma_{in}$ is roughly reciprocal to
the background plasma density $(n_{bg}/\gamma_{\beta}n_0)$.
Considering the initial equilibrium condition
$B^2_0/8\pi=2n_0T$
and the inflow plasma number density
$\gamma_{in}n_{in}\sim n_{bg}$,
some algebra yields
\begin{equation}
\label{eq:sigma2}
\sigma_{in}
\sim \frac{2n_0T}{\gamma^2_{in}n_{in}mc^2}
\sim \frac{2}{\gamma_{\beta}\gamma_{in}}
\frac{\gamma_{\beta}n_0}{n_{bg}}
\frac{T}{mc^2}
\end{equation}

In the case of nonrelativistic antiparallel reconnection,
in the limit of $(v_{in}/v_{out})=\nu \ll 1$
these conditions lead to
\begin{eqnarray}
d_{in} L_{in} v_{in} &=& d_{out} L_{out} v_{out}\\
\Big( \frac{B_{in}^2}{4\pi} \Big) L_{in} v_{in}
&\sim& \Big( 2d_{out}\frac{mv_{out}^2}{2} \Big) L_{out} v_{out}.
\end{eqnarray}
Note that we use the simulation frame density $d=\gamma n$ here.
We obtain a familiar estimate of
\begin{eqnarray}
v_{out} \sim \sqrt{\frac{B_{in}^2}{4\pi m d_{in}}} = \sqrt{2}~V_{A,in}.
\end{eqnarray}
Notice that the Alfv\'{e}n speed in an electron positron plasma
is $V_A=B/[4\pi (2 m d)]^{1/2}$ in the nonrelativistic limit.

Let us consider the relativistic case.
For simplicity,
we assume that incoming flow is flux dominated:
the magnetic energy carries most of the incoming energy.
So, we drop $w_{in}$ from
the left-hand side of equation \ref{eq:eflow}.
This approximation is appropriate for our low-background density runs,
because we set low density background plasmas ($n_{bg}/[\gamma_{\beta}n_0]\ll 1$)
in a nonrelativistic temperature ($T_{bg}\ll mc^2$).
In addition, we approximate the outflow enthalpy term
in the relativistic pressure limit,
\begin{eqnarray}
\label{eq:4p}
w_{out} = ( e_{out} + p_{out} )
\sim \Big( n_{out}mc^2 + \frac{\Gamma p_{out}}{\Gamma-1} \Big) \sim 4 p_{out}
\end{eqnarray}
where $\Gamma\sim 4/3$ is the adiabatic index.
In the outflow region,
the pressure balance across the current sheet
\citep{lyu05} yields
\begin{eqnarray}
\label{eq:pb}
\frac{B_{in}^2}{8\pi} = 2 p_{out}
\end{eqnarray}
Using equations \ref{eq:eflow2} and \ref{eq:pb} and 
introducing $\nu=v_{in}/v_{out}$,
we obtain
\begin{eqnarray}
\Big( \frac{ 1+\sigma_{in} }{ \sigma_{in} } \Big) 
4 p_{out} L_{in} v_{in}
&\sim&
( 8 \gamma_{out}^2 +  4 \nu^2 )
p_{out} L_{out} v_{out}
\end{eqnarray}
The number density yields
\begin{eqnarray}
\label{eq:dratio}
\frac{\gamma_{out}n_{out}}{\gamma_{in}n_{in}}
=
\frac{L_{in}v_{in}}{L_{out}v_{out}}
\sim
\Big( \frac{ \sigma_{in} }{ 1+\sigma_{in} } \Big) 
(2 \gamma_{out}^2 + \nu^2)
\end{eqnarray}
In the limit of $\nu^2 \ll 1$ and $\sigma_{in} \gg 1$,
we obtain
\begin{eqnarray}
\label{eq:dratio2}
\frac{\gamma_{out}n_{out}}{\gamma_{in}n_{in}} \sim 2 \gamma_{out}^2
\end{eqnarray}
in the simulation frame density,
or $(n_{out}/n_{in}) \sim 2 \gamma_{in}\gamma_{out}$ in the proper density.
This analysis tells us that
the relativistic steady state reconnection is not incompressible
in both the simulation frame and the proper frames.
Previous reconnection models \citep{bf94b,lyut03a}
employ an incompressible assumption in the proper frames.
Our analysis shows that
the relativistic reconnection model should be
re-constructed by taking compressibility into account.
Incompressible models are helpful
when and only when $\gamma_{out} \sim 1$.


We can confirm the above argument
by using \citet{lyu05}'s analysis.
For two fluids, the $x$-momentum conservation along the outflow line yields
\begin{eqnarray}
\label{eq:xmom}
2\frac{\partial}{\partial x}
\Big(\frac{\gamma^2_{out}w_{out}v^2_{out}}{c^2}+p_{out}\Big)
\sim \frac{1}{c} J_y B_{out}
\end{eqnarray}
Substituting $(\partial /\partial x) \sim 1/L_{in}$ and
$J_y \sim c{B_{in}}/{(4\pi L_{out})}$ and 
using equation \ref{eq:4p}, we obtain
\begin{eqnarray}
\label{eq:lyu05}
\Big( \frac{4 \gamma^2_{out}v_{out}^2}{c^2} +1 \Big) p_{out}
\sim \frac{L_{in}}{2}\frac{B_{in}}{4\pi L_{out}} B_{out}
\end{eqnarray}
Combining equations \ref{eq:continuity}, \ref{eq:pb}, and \ref{eq:lyu05},
we obtain
\begin{eqnarray}
\label{eq:dratio3}
\frac{\gamma_{out}n_{out}}{\gamma_{in}n_{in}}
=
\frac{L_{in}v_{in}}{L_{out}v_{out}}
\sim \frac{4 \gamma^2_{out} (v_{out}/c)^2+1}{2}
\sim 2 \gamma_{out}^2
\end{eqnarray}

Next, let us consider the current sheet thickness.
In the steady state, the current in the central current sheet
satisfies
\begin{eqnarray}
\frac{c B_{in}}{4\pi L_{out}} = 2q d_{cs} v_y,
\end{eqnarray}
where $d_{cs}$ is the typical number density in the current sheet.
Also, the initial Harris condition satisfies
\begin{eqnarray}
\frac{c B_{0}}{4\pi L} \sim 
\frac{c B_{in}}{4\pi L} = 2q (\gamma_{\beta}n_0) \beta c .
\end{eqnarray}
The ratio of the current sheet thickness yields
\begin{eqnarray}
\label{eq:LL}
\frac{L_{out}}{L}
= \frac{(\gamma_{\beta}n_0) \beta c }{ d_{cs} v_y }.
\end{eqnarray}
We can evaluate the right-hand side
by considering the inflow number density $n_{bg}\sim\gamma_{in}n_{in}$,
a central current sheet density
$\gamma_{in}n_{in} \le d_{cs} \le \gamma_{out}n_{out}$,
and the compressible factor of equation \ref{eq:dratio2},
\begin{eqnarray}
\label{eq:den}
\frac{(\gamma_{\beta}n_0) }{ n_{bg} } \frac{(\gamma_{in}n_{in}) \beta c }{ d_{cs} v_y }
\gtrsim
\frac{(\gamma_{\beta}n_0) }{ n_{bg} }
\frac{ \gamma_{in}n_{in} }{ \gamma_{out} n_{out} }
\frac{\beta c }{ v_y }
\sim
\frac{(\gamma_{\beta}n_0) }{ n_{bg} }
\frac{1}{2\gamma^2_{out}}
\frac{\beta c }{ v_y }.
\end{eqnarray}
Then, equation \ref{eq:LL} can be rewritten as
\begin{eqnarray}
\frac{L_{out}}{L} \gtrsim
\frac{(\gamma_{\beta}n_0) }{ n_{bg} }
\frac{\beta}{2\gamma^2_{out}}
\Big(\frac{ c }{ v_y }\Big).
\end{eqnarray}
We recognize that
the following condition is sufficient for the current sheet expansion,
\begin{eqnarray}
\label{eq:crit}
D=\frac{\gamma_{\beta}n_0 }{ n_{bg} }
\frac{\beta}{2\gamma^2_{out}} \gtrsim 1,
\end{eqnarray}
where $D$ is a discriminant term.
Note that $D$ also indicates the minimum relative thickness
of the steady state current sheet, $L_{out}\gtrsim DL$.

Since we do not have a perfect theory on $\gamma_{out}$,
we refer to the simulation results.
In the case of run 1 ($t/\tau_c=70$),
the plasma $x$-velocity $v_{out}$ and
the electromagnetic field do not satisfy
the frozen condition inside $x \lesssim 10 L$,
and so we regard the region as the diffusion region.
Inside the diffusion region,
the Lorentz factor of the averaged flow
$\gamma^2 = [ 1- (\langle v_{x} \rangle^2 + \langle v_{y}\rangle^2)/c^2 ]^{-1} \sim 2$
stays constant in the diffusion region.
The density stays small $d_{cs} \gtrsim 2 n_{bg}$ and
it gradually goes up in the outflow direction.
Taking the results into account
(e.g. $2\gamma_{out}^2\sim 4$),
we find that the condition (eq. \ref{eq:crit}) is
already satisfied, $D = 1.5>1$.
The observed thickness $\sim 1.9L$ is larger than $1.5L$.
For comparison, we carried out
the other run with dense background plasmas
(run 2; $n_{bg}/\gamma_{\beta}n_0=20\%$), too.
In this case, the current sheet expansion is less significant
(thin line in Fig. \ref{fig:time}\textit{a}).
Although the observed thickness $\sim 1.5L$ is rather larger than $D\sim 0.6$
primary due to the slow $y$-velocity of $\langle v_y \rangle \sim 0.5c$,
it is reasonable that
the current sheet expansion is less significant,
because the inflow delivers more current carriers than in run 1.


\section{GUIDE FIELD CASE}

\begin{figure}[htbp]

\begin{center}
\ifjournal
\includegraphics[width={\columnwidth},clip]{f5.eps}
\else
\includegraphics[width={\columnwidth},clip]{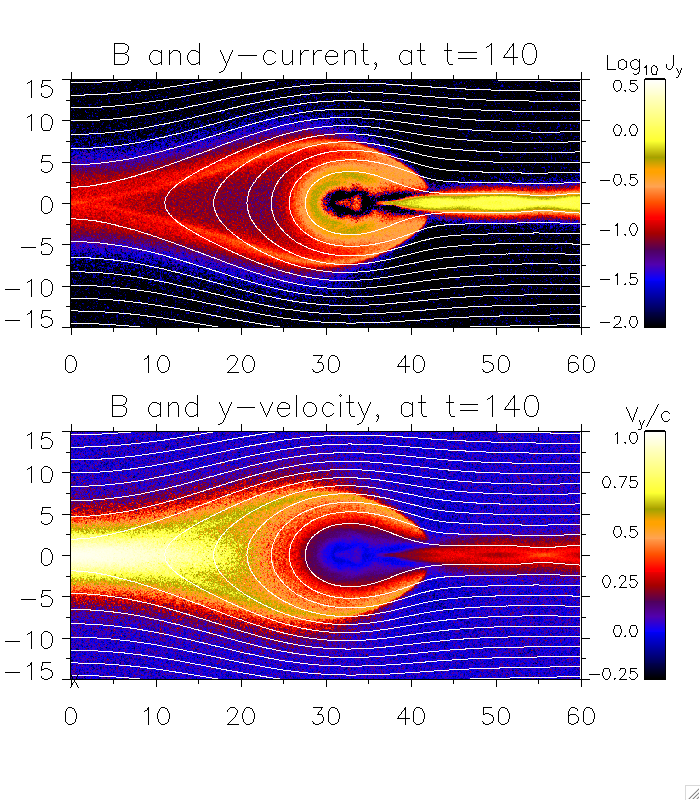}
\fi
\caption{Magnetic field lines and \textit{y}-component of
(\textit{a}) the current density and
(\textit{b}) the average velocity at $t/\tau_c=140$ in run 3.
\label{fig:snapG}}
\end{center}

\end{figure}

\begin{figure}[htbp]
\begin{center}
\ifjournal
\includegraphics[width={\columnwidth},clip]{f6.eps}
\else
\includegraphics[width={\columnwidth},clip]{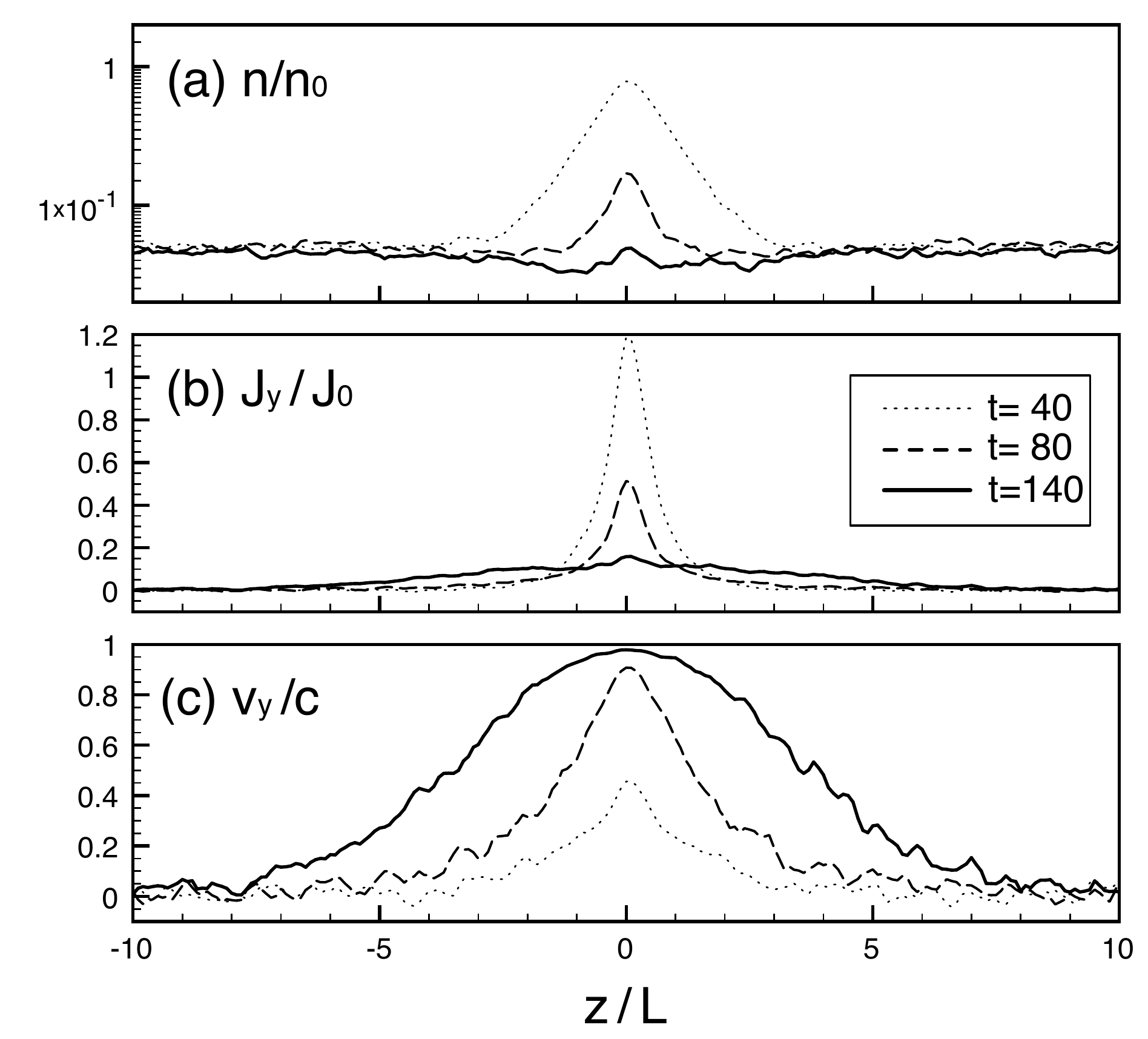}
\fi
\caption{Physical properties along the inflow region
at $x/L=0$
(\textit{a}) normalized plasma density,
(\textit{b}) $y$-component of the electric current, and
(\textit{c}) $y$-component of the plasma average velocity.
\label{fig:plotG}}
\end{center}

\end{figure}

We also study the current sheet problem
in the guide field case.
Snapshots of run 3 with strong guide field $B_G/B_0=1.5$
are presented in Figure \ref{fig:snapG}.
They are at the well-developed stage of $t/\tau_c=140$.
Note that the guide field reconnection evolves
slower than the anti-paralell reconnection.
The typical outflow speed is $v_{out}\sim0.3c$.
The structure of the reconnection region looks
more complicated than the anti-parallel case.
For example, we can see inclined current layers
along the separatrix (Fig. \ref{fig:snapG}\textit{a}).
There are also in-plane electric fields ($E_x,E_z$)
due to the charge separation of positrons and electrons \citep{karl08,zeni08}.
However, in order to discuss
a zeroth-order structure of the diffusion region,
a similar Sweet-Parker approximation would be plausible.
In fact, the $y$-velocity structure,
which is a good indicator of the diffusion region structure,
looks like a rectangular structure (Fig. \ref{fig:snapG}\textit{b}).
Panels in Figure \ref{fig:plotG} present
the plasma properties along the inflow line $x=0$,
similar to those of Figure \ref{fig:plot}.
Surprisingly, the plasma number density in the current sheet is
slightly lower than or almost the same as the inflow density
at the well-developed stage ($t/\tau_c=140$; Fig. \ref{fig:plotG}\textit{a}).
Plasma seems to be incompressible unlike the anti-parallel case.
The plasma average $y$-velocity is even faster:
$\langle v_y \rangle \sim 0.9c$ at $t/\tau_c=80$ and 
$\langle v_y \rangle \sim 0.97c$ at $t/\tau_c=140$ at the center.
This is probably because
the guide field confines particle motion into the $y$-direction.
Importantly, the current sheet looks much thicker than
the anti-parallel case
(Figs. \ref{fig:snapG}\textit{a}, \ref{fig:plotG}\textit{b}, and \ref{fig:plotG}\textit{c}).

Next, let us consider a similar Sweet-Parker model.
We set the guide field in the inflow region to $B_{G,in}$.
The energy flux and pressure balance condition
are modified in the following way,
\begin{eqnarray}
\label{eq:Geflux}
\Big( \frac{B_{in}^2+B_{G,in}^2}{4\pi} \Big)
L_{in} v_{in} &=&
\Big( 2\gamma^2_{out}w_{out}+\frac{B_{out}^2+B_{G,out}^2}{4\pi} \Big)
L_{out} v_{out}\\
\label{eq:Gpb}
\frac{{B_{in}^2+B_{G,in}^2}}{8\pi} &\sim & \Big( \frac{B_{G,out}^2}{8\pi} + 2p_{out} \Big).
\end{eqnarray}
Hereafter, we dropped the $\sigma$-related terms
since $[( 1+\sigma_{in} )/\sigma_{in} ] \sim 1$.
The guide field cases are more flux dominated than the anti-parallel case
due to the guide field magnetic energy
(Table \ref{table}).
The effects of in-plane electric fields are included
in the Poynting flux terms, as guide field contributions.
In addition, flux conservation of the out-of-plane field yields
\begin{eqnarray}
\label{eq:Gbyflux}
B_{G,in}L_{in}v_{in} &=& B_{G,out}L_{out}v_{out}.
\end{eqnarray}
Combining equations \ref{eq:4p}, \ref{eq:Geflux}, and \ref{eq:Gpb},
we obtain
\begin{eqnarray}
\label{eq:Gcrit}
\frac{L_{in}v_{in}}{L_{out}v_{out}}
\sim
\frac{8 \gamma^2_{out} p_{out} + ({B_{out}^2+B_{G,out}^2})/{4\pi}}
{4 p_{out} + ({B_{G,out}^2})/{4\pi}}.
\end{eqnarray}
This immediately tells us that
\begin{eqnarray}
\label{eq:Gcomp}
\frac{\gamma_{out}n_{out}}{\gamma_{in}n_{in}}
\sim
\frac{L_{in}v_{in}}{L_{out}v_{out}}
\gtrsim 1,
\end{eqnarray}
and then equation \ref{eq:Gbyflux} yields
\begin{eqnarray}
B_{G,in}\lesssim B_{G,out}.
\end{eqnarray}
So, the guide field is compressed in the outflow region. 
For guide field larger than the anti-parallel component
$B_{in}\lesssim B_{G,in}$,
more energy is stored in $B_{G,out}$ in the outflow region.
As the guide field becomes even stronger,
the guide field term becomes the largest contributor in equation \ref{eq:Gcrit}
and then equation \ref{eq:Gcomp} becomes closer to the unity.
This means that the guide field brings incompressibility
in the simulation frame.
In the strong guide field limit,
we employ an incompressible condition
\begin{eqnarray}
\gamma_{in}n_{in}\sim d_{cs} \sim \gamma_{out}n_{out}
\end{eqnarray}
and then we obtain a condition similar to equation \ref{eq:crit}
for the strong guide field limit
\begin{eqnarray}
\label{eq:crit2}
D_G = \frac{(\gamma_{\beta}n_0)}{n_{bg}}{\beta} \gtrsim 1,
\end{eqnarray}
where $D_G$ is the discriminant term or the minimum relative thickness. 
It is significant that $D_G$ only depends on the initial configuration parameters,
while the antiparallel counterpart $D$ (eq. \ref{eq:crit})
contains a variable $\gamma_{out}$,
which comes from the plasma compressibility effect (eq. \ref{eq:LL}).
As a result,
the sufficient condition is more easily satisfied than in the anti-parallel case,
so that the current sheet expansion is more significant.

The temporal evolution of the current layer thickness
in the guide field runs is presented
by the dashed line in Figure \ref{fig:time}\textit{a}.
In the guide field cases,
time is re-arranged by some offsets ($\Delta t = 60\tau_c$) due to
the late onset of reconnection,
so that we can directly compare their thickness
in the expanding phase.
In the guide field cases,
the current sheet originally becomes thinner
during the long preonset stage of reconnection.
In the case of run 3,
the outflow guide field is $B_{G,out} \sim 1.7 B_0$
and so the system seems to be sufficiently incompressible.
The sufficient condition (eq. \ref{eq:crit2}) yields $D_G = 6 > 1$,
which predicts that the current sheet expands up to a factor of $6$
in the strong guide field limit.
Considering the weak compressibility,
this is consistent with
the observed thickness of $4L$-$5L$ (Fig. \ref{fig:time}). 
%
In the case of run 4,
the current sheet expands slowly
(dotted line in Fig. \ref{fig:time}\textit{a}),
due to the slowest evolution of reconnection.
The condition is $D_G=1.5 > 1$ in the guide field limit,
which is roughly consistent with the asymptotic thickness
of the current sheet, $\sim 1.3L$.
The parameter $D_G$ may be a good approximation
of the thickness in the guide field cases.
One reason is that plasmas are nearly incompressible.
The other reason is that the plasma $y$-velocity is closer to the light speed
$\langle v_y \rangle \sim 0.9c$,
because their motions are threaded by the guide field.

\section{DISCUSSION AND SUMMARY}

We demonstrated that
the relativistic reconnecting current sheet
self-regulates its thickness, so that
it facilitates the fast reconnection rate.
In order to study the aspect ratio of the diffusion region,
much attention has been paid to the length of the diffusion region.
However, it is found that the relativistic reconnecting layer
changes the aspect ratio by expanding its thickness,
and that we obtain a higher reconnection rate.
\citet{zeni01} briefly reported the current sheet expansion and
argued that the relativistic mass increases the meandering width. 
This explains some aspects, but the situation is more complicated.
Indeed, due to the enhanced reconnection electric field $E_y$,
some incoming plasmas start
(relativistic) meandering motion around the neutral plane ($z=0$),
while others are directly driven to the $\pm y$-directions
without reaching the neutral plane.
The current sheet expansion arises from
the net effect that contains all particle motions there.
\citet{bessho07} reported the faster reconnection rate
in the lower density background condition.
That trend can be consistently explained by our theory.
Based on a relativistic Sweet-Parker analysis,
we developed physical interpretations
and derived some critical conditions.
It is found that the current sheet is likely to expand
when the reconnection inflow is flux dominated, and
it further expands when the guide field is strong.
Although the guide field effect is unclear
in the dense background runs (runs 2 and 4),
low-density cases exhibited a significant difference (runs 1 and 3).
In other words,
in the high-$\sigma$ (low-density) limit of astrophysical interest,
the guide field may play a crucial role in the reconnection topology
by expanding the current sheet.

We also found that
the relativistic Sweet-Parker model should be treated compressibly.
In fact, our simulation snapshot shows
the density ratio of $\gtrsim 2$ in the steady state 
(e.g. Fig. \ref{fig:plot}\textit{a}).
Previous authors
assumed plasma incompressibility \citep{bf94b,lyutikov03},
but
these models should be re-constructed
under the compressible condition,
especially when the outflow Lorentz factor can be large
($\gamma_{out} \gg 1$),
unless the guide field is strong enough. 
The relativistic Petschek models \citep{bf94b,lyu05,tol07} are
out of the scope of the present study,
but our analysis on the guide field reconnection supports
\citet{lyu05}'s argument that
the guide field changes the dynamics of reconnection.
In our case, the guide field introduces the incompressibility,
and then the current sheet is more likely to self-regulate.

We find another new result regarding the outflow speed. 
In the nonrelativistic regime,
the reconnection outflow speed is often
approximated by the inflow Alfv\'{e}n speed.
If this can be applied to the relativistic cases
$\gamma^2_{out} \sim \gamma^2_{A,in} \sim (1+\sigma_{in}) \sim \sigma_{in}$,
equation \ref{eq:sigma2} implies that
the critical condition (eq. \ref{eq:crit}) is 
insensitive to $(\gamma_{\beta}n_0)/n_{bg}$ or $\sigma_{in}$.
However, in our simulations,
the obtained outflow Lorentz factor $\gamma_{out}$
is substantially slower than
that of the inflow Alfv\'{e}n speed ($\gamma_{A,in}$).
The outflow velocity is insensitive or weakly sensitive to
the inflow flux $\sigma_{in}$,
and therefore, the current sheet expansion
does depend on $(\gamma_{\beta}n_0)/n_{bg}$.
It is reasonable that
the current sheet expansion is more significant
when reconnection inflow is more flux dominated,
because it contains fewer current carriers. 
There are several reasons why reconnection outflow is ``slow.''
One reason is that reconnection is unsteady.
The reconnection jet slows down
because it pushes away the preexisting plasmas
at the outflow region in unsteady simulations.
This will be checked by the PIC simulations
with open boundary conditions.
Another is the relativistic pressure effect.
In the relativistic regime,
a relativistically high pressure increases
the plasma enthalpy (eq. \ref{eq:4p}),
which works as an effective inertia,
and then bulk speed becomes slower. 
In order to maintain the current sheet against
the strong magnetic pressure (eq. \ref{eq:pb}),
a relativistically high pressure is required.
We can see this from a calculation which neglects pressure effects.
Without the pressure terms,
the energy and number conservation
(eqs. \ref{eq:continuity} and \ref{eq:eflow})
read
\begin{eqnarray}
( 1+\sigma_{in} )
{2\gamma^2_{in} w_{in}}
{L_{in} v_{in}}
& \sim &
{2\gamma^2_{out} w_{out}}
{L_{out} v_{out}}
\\
\frac{2\gamma^2_{out} w_{out}}{2\gamma^2_{in} w_{in}}
\sim
\frac{\gamma^2_{out} n_{out}}{\gamma^2_{in} n_{in}}
& = & 
\frac{\gamma_{out}}{\gamma_{in}}
\frac{L_{in} v_{in}}{L_{out} v_{out}}
\end{eqnarray}
From the equations, we would obtain an unrealistic super-Alfv\'{e}nic outflow,
$\gamma_{out} \sim ( 1+\sigma_{in} )\gamma_{in} \gg \gamma_{A,in}$
\citep{lyutikov03}.
As a result, we conclude that the pressure effect is critical.





Compared with the one-fluid approximation,
our model deals with both electron fluid and positron fluid independently.
Therefore, the number/energy budget conditions are more reliable,
and we can also discuss the $y$-velocity $|v_y|<c$.
Instead, the difficulty is that
several parameters ($d_{cs}$, $\gamma_{out}$, $v_{out}$ and $v_{y}$)
still have freedom in the current sheet or in the outflow region.
The electron/positron fluids with high Lorentz factor $\gamma_{out}$
may flow toward the $x$-direction or toward the $\pm y$-directions.
The reasonable criteria $(v_y^2+v_{out}^2)<c^2$ in the outflow region
will improve the model. 

It is reasonable that the conditions
(eqs. \ref{eq:crit} and \ref{eq:crit2}) are reciprocal
to the drift speed parameter $\beta$.
When the initial drift speed is fast,
the current sheet is relatively thin,
so that it is more likely to expand.
However, in the relativistic speed limit of $\beta \rightarrow 1$,
the current sheet is likely to expand before
the large-scale reconnection evolves
due to the various instabilities
(e.g. the tearing instability or the drift kink instability).
In the vacuum inflow limit of $n_{bg} \rightarrow 0$,
eventually magnetic reconnection will disappear
because there are no plasmas to carry the steady current.
The induced electric field simply travels away
as an electromagnetic wave, without accelerating plasmas. 
We do not know whether the transition is smooth or drastic,
from the self-regulated current sheet
to the quiet vacuum condition. 
The self-regulation of the current sheet will be found
in relativistic reconnection in an ion-electron plasma, too.
When the electron drift speed increases to the order of $c$,
the electron current layer will similarly expand,
while it is unclear that
the enhanced electric field accelerates ions.
When the broadened electron layer becomes
comparable to the ion layer,
ions and electrons more effectively interact with each other,
and then it may affect the reconnection structure once again.




In summary,
we demonstrated that
the relativistic reconnecting current sheet
self-regulates its thickness,
in order to carry the required current.
A simple Sweet-Parker analysis shows that  
the current sheet expansion is dominant
in the flux-dominated inflow
and that the guide field enhances
the current sheet expansion.
It is also noteworthy that
the relativistic Sweet-Parker reconnection is compressible
and that it is incompressible only in the limit of the strong guide field.

\begin{table}[htbp]
\caption{List of Simulation runs}\label{table}
\begin{tabular}{ccccc}
Run &
1 &
2 & 
3 &
4 \\
\hline
$T/mc^2$ &
1 &
1 & 
1 & 
1 \\
$B_{G}/B_{0}$ &
0 &
0 & 
1.5 & 
1.5 \\
$n_{bg}/\gamma_{\beta}n_0$ &
0.05 &
0.2 & 
0.05 &
0.2 \\
$\sigma_{in}$ &
$\sim 36$ &
$\sim 9$ &
$\sim 115$ &
$\sim 30$ \\
$v_{out}/c$ &
$\sim 0.7$ &
$\sim 0.4$ &
$\sim 0.3$ &
$\sim 0.3$ \\
$\gamma^2_{out}$ &
$\sim 2$ &
$\sim 1.3$ &
$\sim 3$-$4$ &
$\sim 1.2$ \\
\end{tabular}
\end{table}

\begin{acknowledgments}
The authors are grateful to
A. Klimas, M. Kuznetsova and H. Takahashi for useful discussions.
The authors also thank the anonymous referee for
his/her constructive comments and careful evaluation of this manuscript.
This research was supported by JAXA/ISAS,
the NASA Center for Computational Sciences, and
NASA's \textit{MMS} SMART mission.
One of the authors (S. Z.) gratefully acknowledges
support from NASA's postdoctoral program.
\end{acknowledgments}

\end{document}